# Raman scattering investigation of the pressure induced structural phase transition in $LaCrO_3$

V. S. Bhadram[1,2,] *, Abhijit Sen[1], A. Sundaresan[1], and Chandrabhas Narayana[1,] *

[1]*Chemistry and Physics of Materials Unit, Jawaharlal Nehru Center for Advanced Scientific Research, Jakkur P.O., Bangalore 560064 India*

[2]*IMPMC, Sorbonne Université, CNRS, MNHN, 4, place Jussieu, 75005 Paris, France*

We report the pressure dependence of perovskite distortions in rare-earth (*R*) orthochromites ($RCrO_3$) probed using Raman scattering in order to investigate the origin of structural transition from orthorhombic *Pnma* to rhombohedral $R\bar{3}c$ phase in $LaCrO_3$. The pressure induced changes in octahedral tilt modes demonstrates that tilt distortions are suppressed in $LaCrO_3$ and are enhanced in the remaining members of $RCrO_3$ family. This crossover between the two opposite pressure behaviours occurs at a critical *R*-ion radius ≈1.20 Å. We attempted to establish the relation between this unusual crossover and compressibility at Cr- and *R*-sites by probing Raman phonon modes sensitive to the mean bond strength of Cr–O and *R*–O respectively. Finally, we study the bond-length splitting of both $CrO_6$ and $RO_{12}$ polyhedra to ascertain the role of polyhedral self distortion in determining the pressure dependant evolution of perovskite distortions.

## I. INTRODUCTION

Pressure effects on the $ABO_3$ perovskite structure has been an active area of research with several recent publications aiming at the understanding of the pressure evolution of elementary distortions and phase transitions in them.[1-9] The interest is mainly because of the diverse and intriguing properties these materials offer and hydrostatic pressure is an excellent tool that alters the interatomic distances and, consequently, interactions to tune these properties.[4,10-12] In general, the structure of $ABO_3$ perovskites can be obtained from two distortions of the ideal $Pm\bar{3}m$ cubic phase. i) inphase and anti-phase $BO_6$ octahedral rotation (tilts) about the [010] and [101] pseudocubic directions[13] and ii) A-cation displacement along the z pseudocubic direction. The octahedral rotations are the primary order parameter that drives the structure to lower symmetry and the amplitude of the distortion exhibits dependency on the size of A-cation.

Now, what happen to these octahedral tilting distortions under external pressure? There were several rules proposed based on theoretical calculations and experimental results.[1,2,6,8,14,15] Initial reports[1,14-16] postulated that the pressure dependence of $ABO_3$ perovskite structure is based on the octahedral rotations and ratio of compressibilities ($M_A$/ $M_B$) of $AO_{12}$ and $BO_6$ polyhedra respectively. If $AO_{12}$ is more compressible than $BO_6$, i.e., $M_A/M_B > 1$, a transition to a higher symmetry phase occurs and otherwise, tilts continue to enhance with pressure. Later Zhao et al[8] proposed that pressure induced changes in the octahedral distortions mainly depend on the electronic configuration of A, B cations. When the formal charge on the cations is +3 (3:3 perovskites), $AO_{12}$ is more compressible than $BO_6$ leading to decrease in octahedral tilts and, consequently, transition to a higher symmetry structure. However, in our own study,[5] we found that orthorhombic $R^{3+}Cr^{3+}O_3$ (*R* = Lu, Tb, Gd, Eu, Sm) do not follow this rule and the tilt distortions enhance with pressure for smaller *R*-cations and they may be suppressed under pressure for larger *R*-cations. Here, for the first time Raman scattering technique was employed to probe the evolution of tilt distortions under pressure using phonons modes that are sensitive to octahedral tilts. This study is in concurrence with the earlier observation of high-pressure phase transition from orthorhombic *Pnma* to rhombohedral $R\bar{3}c$ in $LaCrO_3$ which is mediated through the suppression of tilting distortions.[12] This phase transition is a unique characteristic of $LaCrO_3$ and observed even at high temperature at ambient pressure.[17] $SmNiO_3$ is another example of *Pnma* system where similar high-pressure phase transition was previously reported.[18]

More recently, Xiang et al,[6] from their first-principles calculations, proposed a set of rules that govern the tilt evolution under pressure. In this report, they showed that both in-phase and antiphase tilts of all the $RCrO_3$ members suppress with pressure which is contrary to our earlier observations.[5] In another recent study by Zhou[3] the results from high-pressure x-ray diffraction on four families of orthorhombic $RMO_3$ (M = Ti, Cr, Mn, Fe) perovskites are incompatible with the existing rules. In this study, the $RCrO_3$ family found to exhibit *R*-ion size dependency in the pressure induced evolution of tilt distortions which is consistent with our report.[5] However, the crossover between the opposite pressure behaviours of $RCrO_3$ members and the origin of pressure induced phase transition observed only in $LaCrO_3$ is still not fully established. Considering the intriguing magnetoelectric multiferroicity[19] and spin-phonon coupling[20] that some $RCrO_3$ compounds exhibit and external pressure being an important alternative to simulate the chemical pressure to tune these properties[5], complete understanding of high-pressure structural evolution of $RCrO_3$ system is warranted.

Here, we are reporting the pressure dependent Raman scattering investigations on three compounds, $NdCrO_3$, $PrCrO_3$ $LaCrO_3$. Together with the support of the data from our earlier report[5] on $RCrO_3$ (*R* = Lu, Tb, Gd, Eu, Sm), we try to establish pressure evolution of structural distortions and the possible crossover event in $RCrO_3$ family. We first present the opposite pressure behaviour exhibited by $LaCrO_3$ as compared to other $RCrO_3$ members through pressure dependence of in-phase and anti-phase tilt modes. We explain this unusual behaviour by comparing the $CrO_6$ and $RO_{12}$ polyhedral compressibilities probed using phonon modes sensitive to Cr-O and *R*-O bond strengths respectively. Finally, to gain further support to our arguments, we compare the intrinsic polyhedral distortion of several members of the $RCrO_3$ family using the ambient and high-pressure crystal structural data available in the literature.[3,21]

## II. EXPERIMENTAL METHODS

Polycrystalline powder samples of $LaCrO_3$, $NdCrO_3$ and $PrCrO_3$ were prepared via solid state route. The details of the sample synthesis are given in Ref. 19. Stoichiometric mixtures of rare earth oxides ($La_2O_3$, $Nd_2O_3$ and $Pr_6O_{11}$) and $Cr_2O_3$ were heated stepwise with several intermittent grinding. At first step, heating was carried out at 1473 K for 24 hours followed by grinding. Second step involved heating the resultant ground mixtures at 1673 K for 24 hours followed by grinding. In the third step the ground mixtures were pelletized and the pellets were heated at 1673 K for 24 hours. X-Ray powder diffraction patterns were recorded in Bruker D-8 advance diffractometer and Panalytical Empyrean diffractometer using Cu source ($\lambda$ = 1.5406 Å). Reitveld refinement of the powder patterns were carried out to determine the phase purity (see Fig.S1 in supplementary Information). Profile peak fitting were carried out using pseudo-Voight function. All three compounds found to be of pure crystalline single phases.

The powder samples were pelletized and were crushed into small pieces to pick the pieces of right size for high pressure experiments. A membrane type diamond anvil cell (DAC) from BETSA, France has been used for high pressure experiments. The diamond culet size is 400 microns. Methanol, ethanol (4:1) mixture has been used as a pressure transmitting medium and for *in situ* pressure measurements a ruby chip has been loaded along with sample inside DAC. Raman spectra of the chromites have been collected using a custom-built Raman set up equipped with a laser with excitation wavelength 532 nm and laser power of ≈ 1 mW (power density: 32 $kW/cm^2$) at the sample inside the DAC.

## III. RESULTS AND DISCUSSION

Pressure dependence of the Raman spectra of $LaCrO_3$ up to 15 GPa is shown in Fig 1a and that of $PrCrO_3$ and $NdCrO_3$ are shown in Fig. S2 (Supplementary Information). For orthorhombic (space group, *Pnma*) systems, group theory predicts 24 Raman active modes. But, experimentally, only 10-12 modes are seen as all the other modes could be too weak to be detected. The low frequency modes (below 200$cm^{-1}$) involve the *R*-ion motion and high frequency modes involve $O^{2-}$ motion. The detailed assignment of these modes is given elsewhere.[20,22] Here we briefly describe the modes that are relevant to the present studies. $A_g(7)$ mode is related to rare-earth and oxygen mixed motion in $RO_{12}$ polyhedra and can be a representative of *R*-O bond strength. $A_g(1)$ mode is originated from the stretching of Cr–O bonds and it appears at the high frequency end of the spectrum. In the mid-frequency range, $A_g(4)$ and $A_g(2)$ are $CrO_6$ octahedral rotational modes and are sensitive to octahedral inphase (angle, $\theta$) and anti-phase (angle, $\varphi$ ) tilts respectively. In fact, clear linear dependence of rotational (tilt) mode frequencies on the corresponding tilt angles was proposed by Iliev et al [23] for orthomanganites, and for rare-earth orthochromites by Weber et al[22] and Bhadram et al.[8] These tilt modes are also called as soft modes as their frequency tend to become zero when tilt angles become zero which means tilt modes are absent in the undistorted perovskite structure.[23]

The pressure dependence of the Raman spectrum of $LaCrO_3$ (shown in Fig.1a) depicts that it undergoes a structural phase transition from orthorhombic (*Pnma*) to rhombohedral ($R\bar{3}c$) at around 4.5 GPa. The transition pressure matches well with the previous neutron diffraction studies on $LaCrO_3$.[12] It is evident from Fig.1a that the transition from *Pnma* to $R\bar{3}c$ results in the complete reconstruction of the Raman spectrum. The spectra are deconvoluted and the pressure dependence of the Raman phonon frequencies is shown in Fig 1b. While the number of observed phonon modes in the *Pnma* phase is 9, it has reduced to 4 in the $R\bar{3}c$ phase. According to group theory, there are 5 Raman active modes ($4E_g$+$1A_g$) predicted for $R\bar{3}c$ phase.[24,25] Among them, three $E_g$ and one $A_g$ modes can be observed in our Raman spectra. The other $E_g$ mode is a lattice mode whose frequency is too low to be detected using our spectrometer. It is to be noted that all the modes in the high pressure $R\bar{3}c$ phase has counterparts in *Pnma* phase too which is because all these modes arise from the cubic zone-boundary modes of ideal perovskite structure (P $m\bar{3}m$).[24-26] Interestingly, the *Pnma* to $R\bar{3}c$ phase transition is absent in the case of $PrCrO_3$, $NdCrO_3$ (see Fig.S1 of supplementary Information) and also in other members of $RCrO_3$.[5]

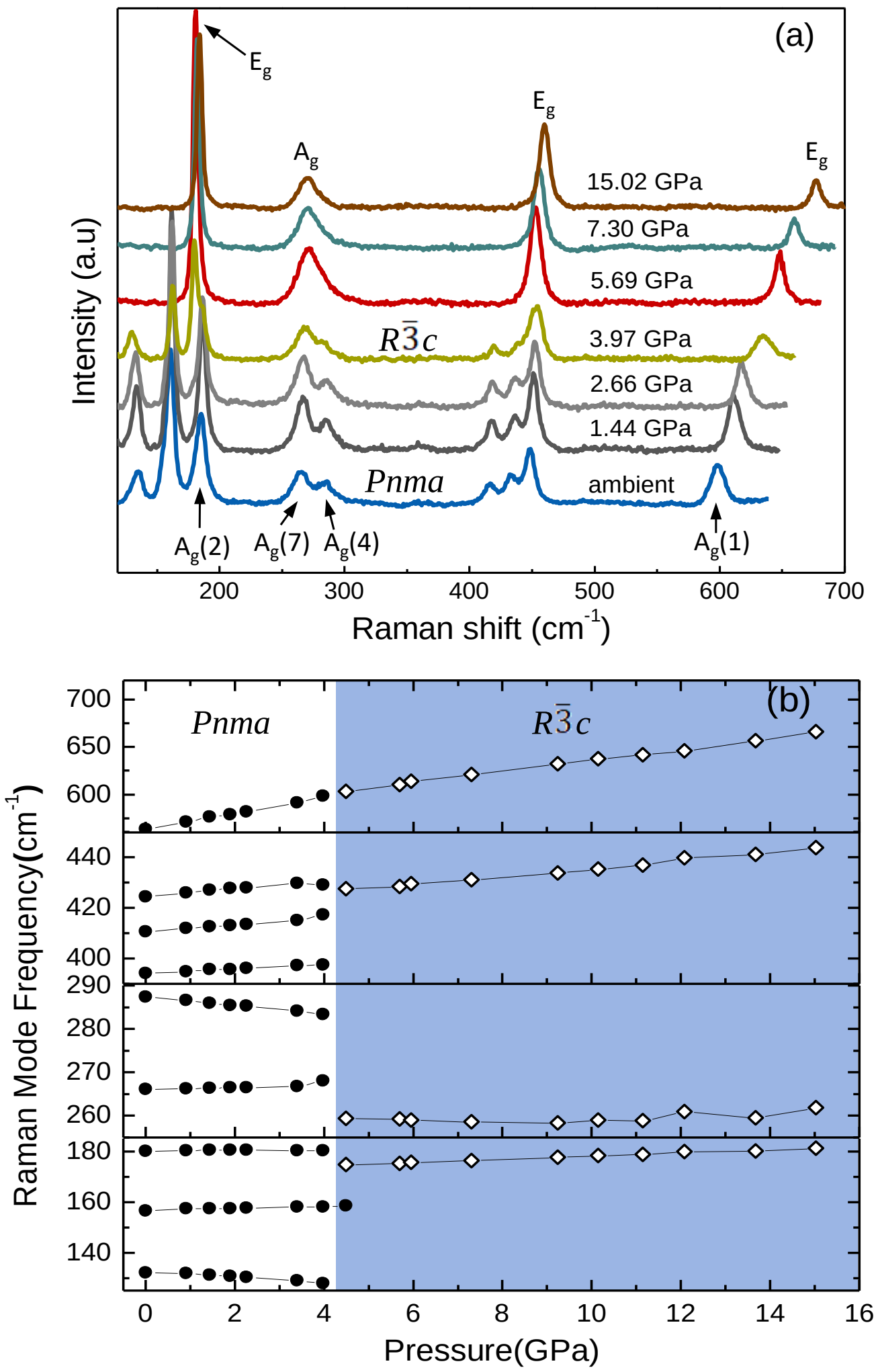

FIG. 1. (a) Raman spectrum of $LaCrO_3$ at different pressures. (b) Phonon frequencies of $LaCrO_3$ as a function of pressure.

In general, the symmetry breaking transitions in $ABO_3$ perovskites can be explained through A-cation displacement and $BO_6$ octahedral tilts.[27] The A-cation displacement is seen in perovskite structures without centrosymmetry. Thus the main component of the structural transition from *Pnma* to $R\bar{3}c$ in $RCrO_3$ is $CrO_6$ octahedral tilts. To probe the pressure dependence of the octahedral tilts using Raman, octahedral rotational (tilt) modes are of significance, thanks to the linear dependence of their frequencies with octahedral tilt angles. [22] In our previous report,[5] we used this dependency to probe the evolution of octahedral tilts under pressure. We clearly showed that in $RCrO_3$ (*R* = Lu, Tb, Gd, Eu, Sm) octahedral tilts increase with pressure at a rate that decreases with *R*-ion radius (IR).

In Fig. 2, pressure dependence of frequencies of tilt modes $A_g(4)$, $A_g(2)$ of $PrCrO_3$, $NdCrO_3$, and $LaCrO_3$ are shown. For ease of comparison, we are also showing our earlier data on $LuCrO_3$ but omitted the other members of $RCrO_3$ family to avoid the cluttering of the data. It is clear from Fig. 3 and the data shown in Ref. 5, within the pressure range for best linear fits, except for $LaCrO_3$, the frequencies of rotational modes of all compounds in $RCrO_3$ series decrease with increase in pressure. This means that octahedral tilts increase under pressure for *R* = Lu to Pr and only in case of $LaCrO_3$, they reduce with increase in pressure. As a consequence of this, only $LaCrO_3$ undergoes pressure induced structural transition to a high-symmetry phase. Also, from Fig 2a, b, we can clearly see a difference in the pressure behaviour of tilt modes among different $RCrO_3$ compounds. To visualize this, we have plotted the rate of change in the rotational mode frequency (dω/dP) against the *R*-ion radius (IR) in Fig. 2c. Here, the data on $PrCrO_3$, $NdCrO_3$, and $LaCrO_3$ are from the present work and rest of the data are taken from Ref.5. The quantity dω/dP is positive and decreases with increase in IR and it becomes negative for $LaCrO_3$. The crossover between positive to negative trend occurs at IR ≈ 1.20 Å which falls between $PrCrO_3$ and $LaCrO_3$. It is noteworthy that similar crossover event occurs in rare-earth orthoferrites ($RFeO_3$) but at much smaller IR (around Eu).[2] Another contrasting feature is that those $RFeO_3$ compounds whose tilts reduce with increase in pressure would display discontinues volume change at P > 40 GPa, but space group remains *Pnma*.

In $ABO_3$ perovskites, apart from tilt angles, tilt mode frequencies also depend on B–O bondlengths as established previously for orthoferrites.[28,29] This relation can be written as follows for $RCrO_3$

$$\omega = (\alpha_1 - \alpha_2 \langle Cr - O \rangle)\emptyset \qquad (1)$$

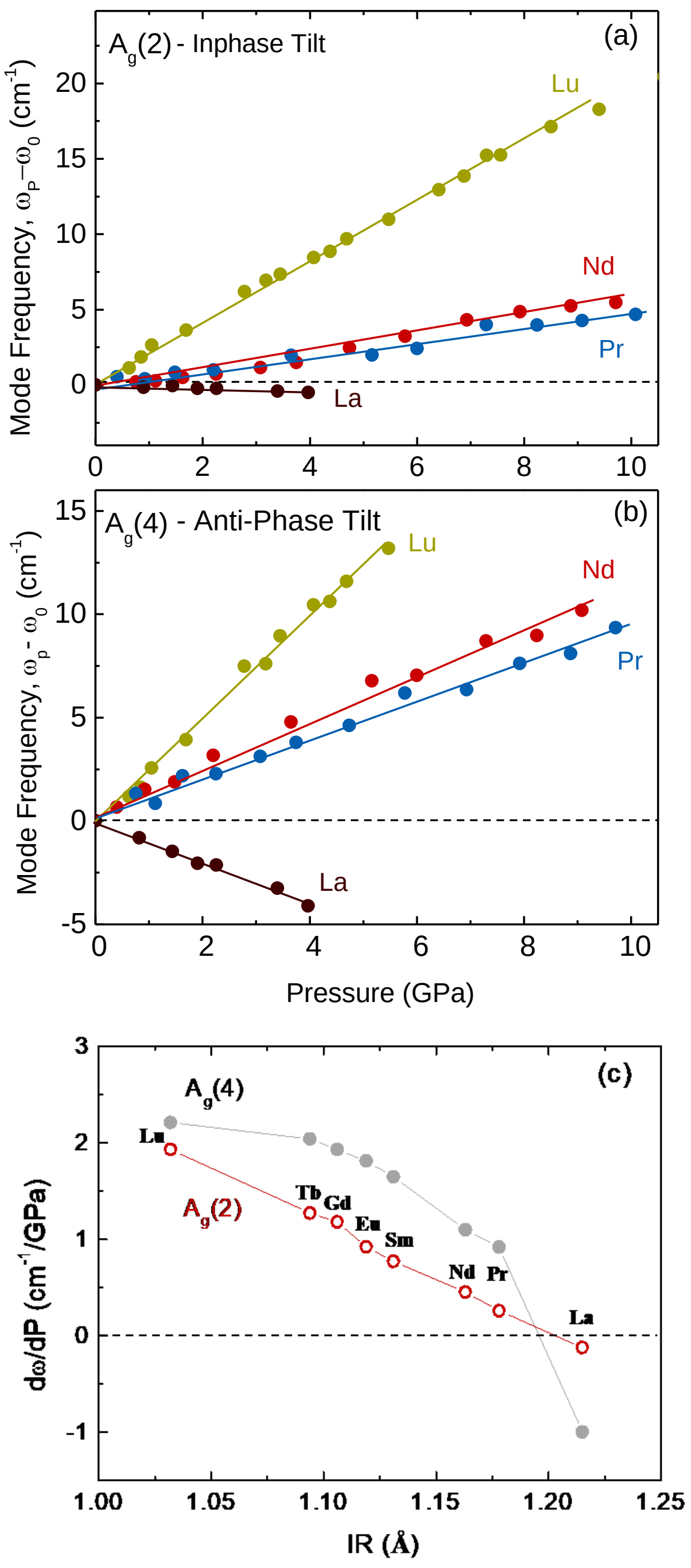

FIG. 2. Pressure dependence of tilt mode frequencies correspond to (a) in-phase tilt (b) antiphase tilt. Solid lines are the best linear fits to the data below 10 GPa pressure. (c) The pressure derivative of tilt mode frequency as a function of rare-earth cation radius (IR).

Here $\alpha_1$ = 109.1 $cm^{-1}$/deg, $\alpha_2$= 42.3 $cm^{-1}$/deg (values are taken from Ref. 28) and $\Phi$ is the tilt angle and $\langle Cr-O\rangle$ is the mean Cr–O bond length. In the above relation the multiplication factor on the right hand side represents the contribution from both actual tilt angle change and isotropic reduction of the $CrO_6$ octahedra volume. From Eq.1, the pressure dependence of tilt mode frequency can be written as

$$\frac{d\omega}{dP} = [\alpha_1 - \alpha_2 \langle Cr-O\rangle(P)]\frac{d\emptyset}{dP} - \alpha_2\emptyset(P)\left(\frac{d\langle Cr-O\rangle}{dP}\right) \qquad (2)$$

We already know dω/dP from Raman measurements, but to estimate the $\frac{d\langle Cr-O\rangle}{dP}$ for different IR we need the values of $\emptyset$, $\frac{d\emptyset}{dP}$ and $\langle Cr-O\rangle$ in the vicinity of arbitrary pressure P. For orthorhombic *Pnma* systems, tilt angles can be deduced from lattice parameters using the equations of Megaw *et al*, [13,30]

$$\theta = \cos^{-1}\left(\frac{a}{b}\right), \varphi = \cos^{-1}\left(\frac{\sqrt{2}a}{c}\right) \qquad (3)$$

Please note that these relations hold good for an orthorhombic system with $a < \frac{c}{\sqrt{2}} < b$ as a consequence of the tilting distortions. This condition is fulfilled for the pressure range of 0-5 GPa for most of the $RCrO_3$ except for $PrCrO_3$ and $LaCrO_3$.[3] For the later systems, $a$, $b$, $\frac{c}{\sqrt{2}}$ cross each other and, thus, Eq (3) is no good for them. Tilt angles of $LuCrO_3$, $GdCrO_3$ and $NdCrO_3$ are estimated from Eq (3) using the lattice parameters in the vicinity of P = 0 GPa as obtained from neutron diffraction measurements published in Ref. 3. We chose P = 0 GPa for our calculation since the values of $\langle Cr-O\rangle$ and $\emptyset$ easily available at that point. The corresponding pressure derivatives are listed in Table 1. By plugging $\frac{d\omega}{dP}$, $\frac{d\emptyset}{dP}$ in Eq (2) , we estimated $\frac{d\langle Cr-O\rangle}{dP}$ for both inphase and antiphase tilts of three compounds as shown in Table 1. The estimated $\frac{d\langle Cr-O\rangle}{dP}$ is varied within 0.001-0.002 Å/GPa when using two different rotational modes. From Table 1, it is clear that when the rare-earth ionic radius increases the pressure rate at which $CrO_6$ compressed also increases. This means, for smaller rare-earth cations, $CrO_6$ octahedra compressed at lower rates and on the contrary $CrO_6$ octahedra compressed at larger rates for larger rare-earth cations.

In general, $CrO_6$ octahedral compression alone cannot determine the fate of octahedral tilts under pressure. The compressions at *R*-site also need to be taken into account.[1,8] But, direct observation of change in bondlength under pressure may be difficult using x-ray diffraction since the peak intensities are highly sensitive to stress which makes it difficult to refine the atomic positions. Although neutron diffraction gives accurate atomic positions and bond length information, currently, such a study is reported only for $LaCrO_3$.[12] In the absence of accurate bond length information, pressure dependence of $A_g(1)$ and $A_g(7)$ phonon mode frequencies can be used to observe the pressure induced compression at Cr- and *R*-sites respectively. These phonon frequencies are directly related to Cr-O and *R*-O bond strengths, respectively.[5,22] In fact, in our previous report[5] on $RCrO_3$ (*R* = Lu, Tb, Gd, Eu, Sm), we have established that the pressure derivatives (dω/dP) of $A_g(1)$ and $A_g(7)$ frequencies represent the rate of volume reduction/compressibility of $CrO_6$ and $RO_{12}$ polyhedra respectively. Large dω/dP indicates highly compressible polyhedra conversely, rigid polyhedra results in small dω/dP value. Here, we are extending the analysis to $LaCrO_3$, $NdCrO_3$ and $PrCrO_3$ in Fig 3a. With the increase in IR, dω/dP of $A_g(1)$ increases monotonously from *R* = Lu to La. This indicates highly compressible $CrO_6$ octahedra for compounds with large IR which is consistent with the behaviour of $\frac{d\langle Cr-O\rangle}{dP}$ discussed above. On the other hand, dω/dP of $A_g(7)$ decreases with increase in IR from Lu to La and approaches zero for the case of $NdCrO_3$, $PrCrO_3$ and $LaCrO_3$. This is a clear indication of greater rigidity of $RO_{12}$ polyhedra in larger *R*-ions as compared to the smaller ones and this change in trend is found to be occurring around IR ≈1.125Å. It is already known that the pressure dependent behaviour of octahedral distortions is controlled by the relative compressibility of both $CrO_6$, $RO_{12}$ polyhedra.[1,8] Thus, in the case of $LaCrO_3$, highly compressible $CrO_6$ octahedra together with almost incompressible $RO_{12}$ leads to the suppression of octahedral tilts and pressure induced phase change from *Pnma* to $R\bar{3}c$.

Table 1. $d\langle Cr-O\rangle\,/\,dP$ calculated from Eq. (2), in the vicinity of P = 0 GPa using either the in-phse or antiphase tilt angles and tilt mode frequencies for different rare-earth chromites.

| System | $d\theta/dP$ (deg/GPa) | $d\varphi/dP$ (deg/GPa) | $d\langle Cr-O\rangle\,/\,dP$ (Å/GPa) | |
|---|---|---|---|---|
| | | | Antiphase tilt (θ) | In-phase tilt (φ) |
| $LuCrO_3$ | 0.07715 | 0.01006 | -0.003958 | -0.002160 |
| $GdCrO_3$ | 0.03720 | 0.01760 | -0.004970 | -0.002921 |
| $NdCrO_3$ | 0.16330 | 0.05798 | -0.009100 | -0.006830 |

To rationalize the systematic change in the compressibility of $CrO_6$ and $RO_{12}$ with varying IR, we looked into the intrinsic bond-length distortions of these polyhedra. Generally, in cubic $ABO_3$ perovskites, all 12 A–O bonds and 6 B–O bonds will have identical bond- lengths. But, the symmetry lowering occurs depending on the cation size and structural instabilities which leads to dissimilar bond-lengths and lowering of A-cation coordination from 12 to either 8 or 6 depending on the crystal symmetry.[31] However, to present more general perspective that can be extended to all perovskites, in the present bon-length distortion analysis in orthorhombic $RCrO_3$, we consider all 12 *R*-O bonds. We define bond length splitting parameter $\rho = \left[\sum_{i=1\to n}(l_i - l_0)^2\right]^{1/2}$ where n is the coordination number and $l_0$ is mean bond length. The values of $\rho_{Cr-O}$ (for $CrO_6$) and $\rho_{R-O}$ (for $RO_{12}$) of several $RCrO_3$ compounds are determined using the refinement results of X-ray diffraction data in Ref. 21 and our present data on three compounds. The estimated $\rho$ values as a function of IR are shown in Fig.3b. For $LaCrO_3$, we are also showing the $\rho$ value determined from neutron diffraction analysis from Ref. 12 For the entire $R$CrO$_3$ series, $\rho_{Cr\text{-}O} << \rho_{R\text{-}O}$ which indicates that the distortion at Cr site is negligible as compared to $R$ site. The variation in $\rho_{Cr\text{-}O}$ with respect to IR is negligible for IR< 1.125 Å. But, for IR< 1.125 Å, there is slight upward trend in $\rho_{Cr\text{-}O}$ is observed. On the other hand, $\rho_{R\text{-}O}$ monotonously decreases with increase in IR from Lu to La and it is lowest for $LaCrO_3$. A clear slope change is observed around IR ≈1.125 Å which is consistent with the behaviour of bond compressibilities (Fig. 3a) and tilt modes (Fig. 2c). This clearly indicates bond length splitting as a function of IR could be helpful in determining the overall effect of pressure in $R$CrO$_3$ and other perovskites.

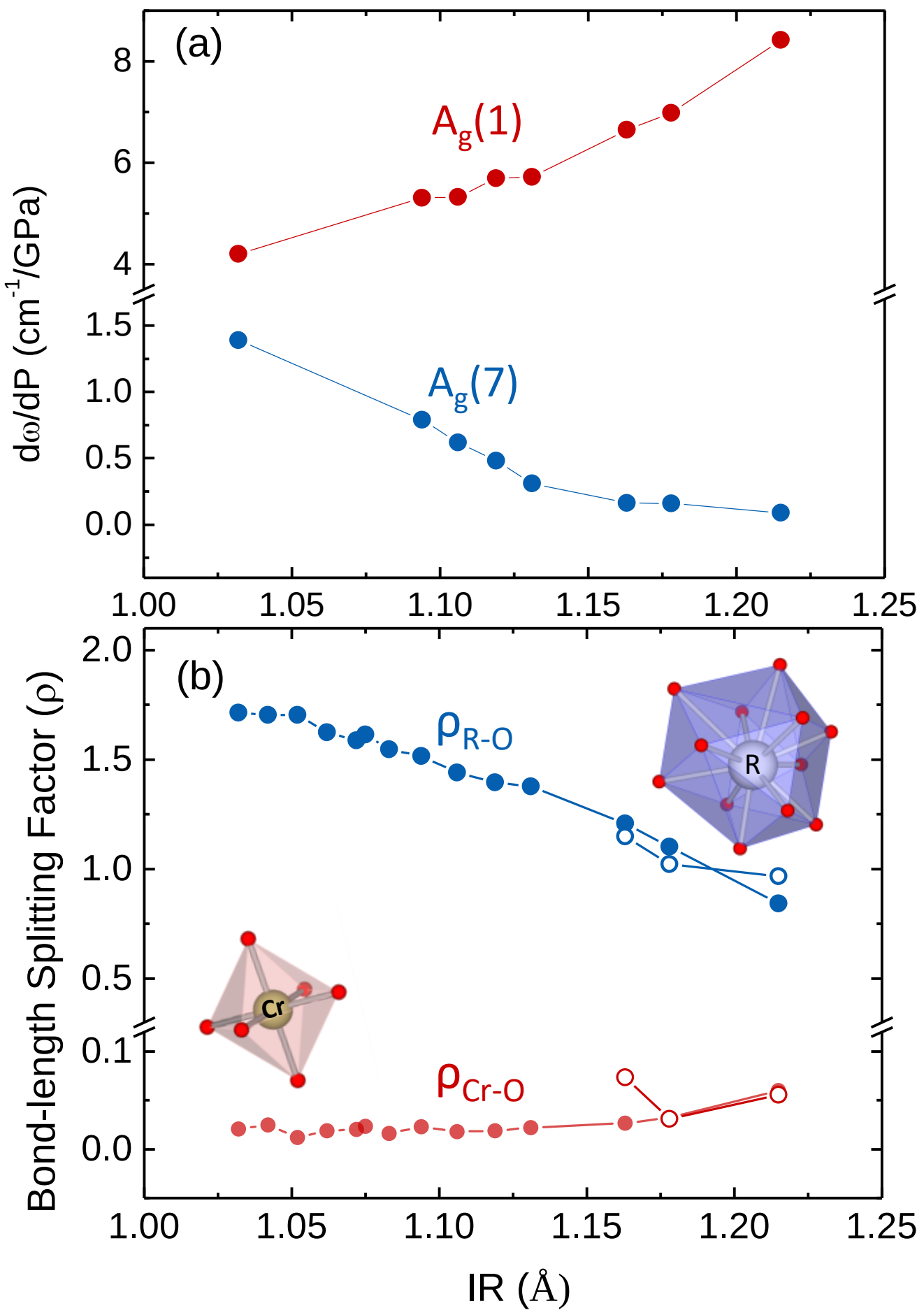

FIG. 3. Rare-earth cation size (IR) dependence of (a) $CrO_6$ and $RO_{12}$ polyhedral mode frequencies which are directly proportional to Cr-O and *R*-O bond strengths, respectively, and consequently, inversely proportional to corresponding compressibilities. (b) Bond-length splitting factors ($\rho_{R-O}$ and $\rho_{Cr-O}$) at ambient pressure estimated using the crystal structure information from X-ray diffraction studies. Solid symbols represents the data published in Ref. 21 and, open symbols corresponds to the present data on $NdCrO_3$, $PrCrO_3$ and $LaCrO_3$.

## IV. SUMMARY

The pressure dependence of orthorhombic distortion in rare-earth orthochromites ($R$CrO$_3$) has been studied using Raman light scattering spectroscopy in order to probe the origin of structural phase transition at 4.5 GPa from orthorhombic to rhombohedral phase observed only in $LaCrO_3$. Based on pressure dependence of $CrO_6$ tilt mode frequencies, we found that $CrO_6$ octahedral tilts increase with pressure but at a rate that decreases with increase in $R$-ion radii (IR). And the crossover between positive and negative rate occurs for the critical IR ≈1.20 Å which falls between $PrCrO_3$ and $LaCrO_3$. The pressure dependant behaviour of the volume reduction (compression) in $CrO_6$, $R$O$_{12}$ polyhedra which drive the tilt distortions is qualitatively presented using Raman modes sensitive to Cr-O, $R$-O bond strengths, respectively. The pressure induced compressions of both polyhedra found to show opposite trends with increase in IR and a slope change is observed at IR ≈ 1.125 Å. Almost negligible compressibility of $LaO_{12}$ together with large compressibility of $CrO_6$ found in $LaCrO_3$ could be the driving force for structural transition in $LaCrO_3$ at high pressures. The unique high-pressure behaviour of $LaCrO_3$ as compared to other members of the $R$CrO$_3$ family was further explained by bond length splitting factors of $CrO_6$ and $R$O$_{12}$ derived from ambient structural data available in the literature. The bond length splitting factors of both polyhedra follow the same trend with respect to IR as their corresponding compressibilities. This suggests that bond length splitting could be a useful parameter to determine the effect of pressure on the perovskite distortions.

* venkata.s.bhadram@gmail.com
* cbhas@jncasr.ac.in

# Supplementary Information for "Raman scattering investigation of the pressure induced structural phase transition in $LaCrO_3$"

V. S. Bhadram[1,2,] *, Abhijit Sen[1], A. Sundaresan[1], and Chandrabhas Narayana[1,] *

[1]*Chemistry and Physics of Materials Unit, Jawaharlal Nehru Center for Advanced Scientific Research, Jakkur P.O., Bangalore 560064 India*

[2]*IMPMC, Sorbonne Université, CNRS, MNHN, 4, place Jussieu, 75005 Paris, France*

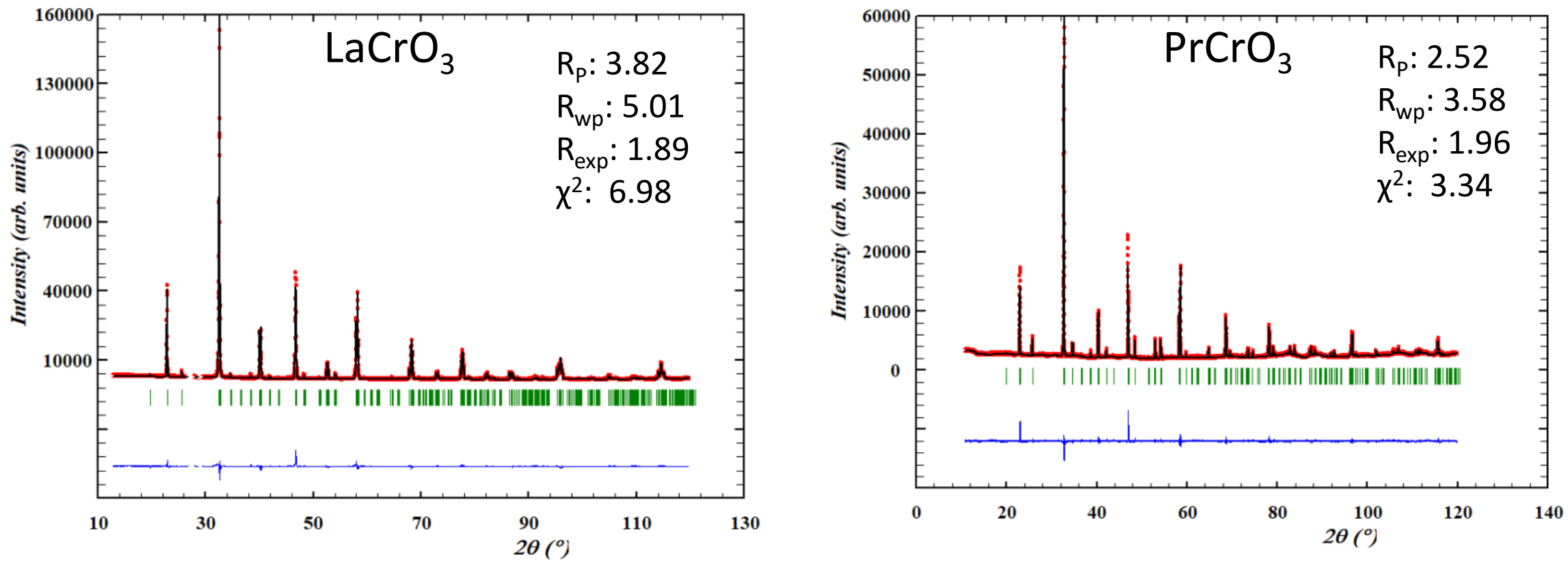

FIG. S1. The Rietveld refinement results of $LaCrO_3$ and $PrCrO_3$. The pattern in red color is the original diffraction; the pattern in black color is the fitting result. Green vertical bars indicate the Bragg peak positions for the *Pnma* phase.

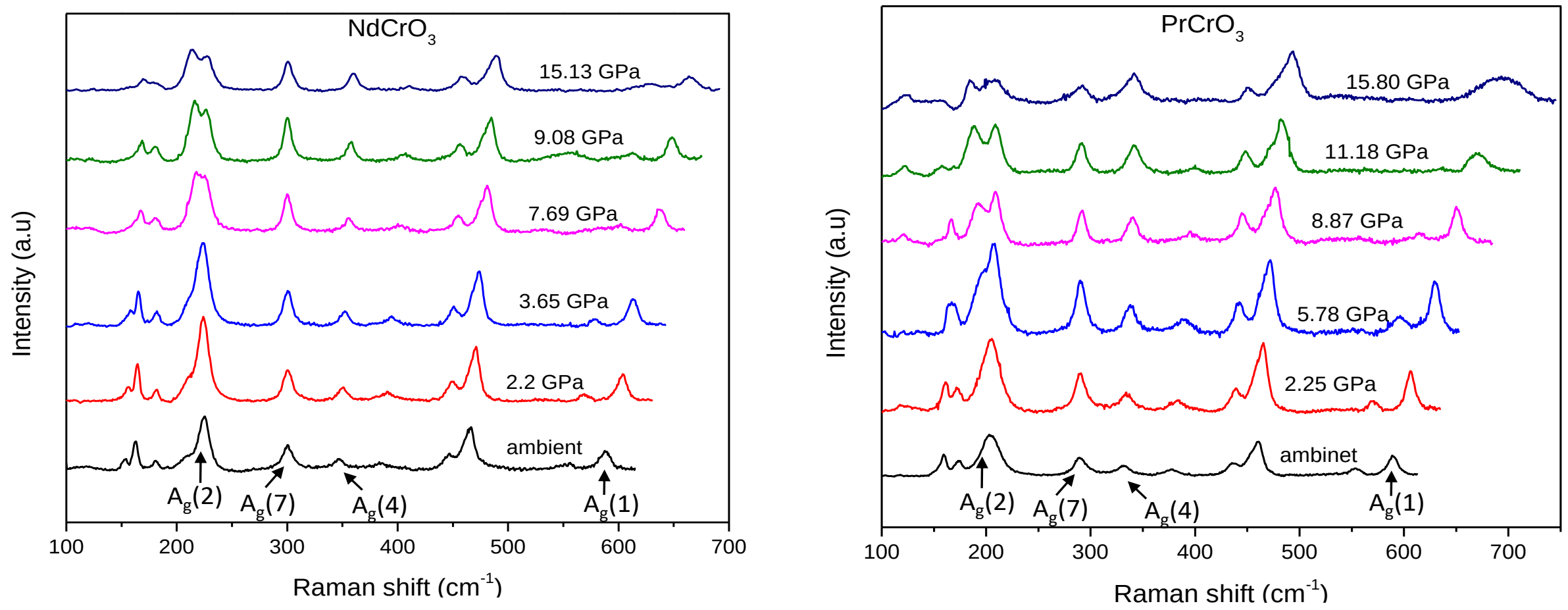

FIG. S2. Raman spectra of $NdCrO_3$ and $PrCrO_3$ as a function of pressure. $CrO_6$ octahedral in-phase and antiphase rotational modes ($A_g(2)$, $A_g(4)$), and polyhedral modes sensitive to Cr-O and *R*-O bonlengths ($A_g(1)$, $A_g(7)$) are indicated.